\documentclass{article}

\usepackage{graphicx}
\newcommand{\be}{\begin{equation}}
\newcommand{\ee}{\end{equation}}
\renewcommand{\k}[2]{\frac{#1}{#2}}

\newcommand{\pcd}[2]{\frac{d^n \! #1}{d #2^n}}

\def\s{\,\,\,\,}
\def\t{\tau}
\def\d{\delta}
\def\q{\Gamma_{d}}
\def\z{\zeta}
\def\ra{\rightarrow}
\def\sumj{\sum_{j=1}^{\t/\d}}
\def\sumjj{\sum_{j=1}^{\infty}}
\def\sumr{\sum_{j=1}^{r/\d}}

\title{\Large \bf Fractals in Small-World Networks With Time Delay}

\author{ Xin-She Yang     \\
        Department of Fuel and Energy and Department of Applied Mathematics \\
        University of Leeds, LEEDS LS2 9JT, UK }

\date{}
\begin{document}
\maketitle

\begin{abstract}

The small-world networks recently introduced by Watts and Strogatz
[Nature 393, 440 (1998)] has  attracted much interests
in studying the interesting properties of the networks without time-delay.
However, a signal or influence travelling on the small-world networks often
associated with time-delay features which are very common in
biological and physical networks. We develop an analytical approach
as well as numerical simulations to try to characterize the effect
of time-delay on the properties of small-world networks. An
analytical expression of the fractal dimension
of the small-world networks is given and thus compared with the results from
numerical simulations. Analysis shows that
small-world networks with time-delay generally have
the multifractals property.

\end{abstract}

\noindent {\bf Citation detail:}, X. S. Yang, Fractals in small-world networks with time delay,
 {\it Chaos, Solitons \& Fractals}, {\bf 13} (2), 215-219 (2002).

\section{Introduction}

The properties of complicated networks such as internet servers,
power grids, forest fires and porous media are mainly
determined by the structure of connections
between the vertices or occupied sites.
These networks can be generally represented as a graph network.
Obviously, one extreme of such a graph is the regular network
which has a high degree of local clustering and the average distance between
the vertices is quite large, while the other extreme
is the random network which shows negligible local clustering and the average
distance is relative small. Recently, Watts and Strogatz [1] have presented
a model for small-world networks with a high degree
of local clustering and a small average distance. Such a
small-world phenomenon can be obtained by adding randomly only
a small fraction  of the connections, and some common networks such
as power grids, film stars networks and neural networks behave
like small-world networks [2-6].

Recently, Moukarzel [3] studied the spreading and shortest paths
in system with sparse long-range connections by using the small-world model.
The spreading of some influence such as a forest fire, an infectious disease
or a particle in percolating media is studied using a simple rule: at each
time step, the influence propagates from the infected site to all uninfected
sites connected to it via a link although this link is not necessary physical.
A long-range connection or shortcut can simulate the spark that starts a new
fire spot, the infect site (say person with flu) suddenly travels to a new
place (site), or a portable computer with virus that start to connect to the
network a new place. The Newman-Watts model [2] and  Moukarzel model
[3] were mainly focusing the immediate response of the network by
adding randomly some long-distance shortcuts; there was no time delay in the
network systems. However, in reality, a spark or an
infection can not start a new fire spot or new infection immediately,
it usually takes some time $\d$, called ignition time or waiting time,
to start a new fire or infection. Thus the existing models are
no longer able  to predict the response in the networks or systems
with time delay.

The purpose of the present work is to investigate the small-world networks
with time delay by using or extending the existing
the small-world theoretical framework developed by Watts and Strogatz (1998)
so as to characterise the fractal dimension of  the small-world network with time-delay.
The present model will
generally lead to a delay differential equation, whose solution is usually
very difficult to obtain if it is not impossible. Thus the numerical
simulation becomes essential [1]. We will take the analytical
analysis as far as possible and compare with the results from numerical
simulations. In addition,
we shall focus mainly on the 1-D and 2-D networks to study  the
effect of time-delay on fractal dimension  and other properties of
the small-world networks.

\section{Delay Differential Equation for the Small-World Networks}

Now considering a randomly connected network with
n-dimensional lattice [7,8,9,10] and $n=1,2,3, ...$.
Assuming an influence spreads with a constant velocity $u$ in all
directions and a new infected spot in the other end of a shortcut will start
but with a time delay $\Delta$. Following the method developed by Newman
and Watts [2] and Moukarzel [3], the total infected
volume $V(t)$ comes from two contributions: one is the primary influented
volume, and the other contribution is the secondary volume by shortcuts.

The primary volume at time $t$ is the influenced part inside a (hyper)
sphere of radiu $u t$ and the surface $\q (ut)^(n-1)$, so the primary
volume is $\q u^n \int_0^{t} \z^{n-1} d\z$. When a primary sphere
meets a shortcut end (with a density $2p$ of shortcuts on the
network where $p$ is the probability of the longe-range shortcuts), a new
secondary sphere starts randomly at the other end of the shortcut. This
secondary volume with a time-delay $\Delta$ is $\q u^n \int_0^{t}
[2p V(t-\z-\Delta)] \z^{n-1} d\z$.  By using a
continuum approach to the network, the total volume satisfies
the time-delay equation of Newman-Watts-Moukarzel type
\be
V(t)=\q u^n \int^{t}_0 \z^{n-1} [1+\xi^{-n} V(t-\z-\Delta)] d\z,
\s n=1,2,3, ... \label{equ-1}
\ee
where $\q$ is shape factor of a hypersphere in n-dimensions. The
length scale of Newman-Watts type [2] can be defined as $\xi=1/(2p)^{1/n}$.
However, by slightly rescaling the volume $V$, a generalized
definition is more appropriate. That is
\be
\xi=\k{1}{(2 p k n)^{1/n}},
\ee
where $k$ being some fixed range. By proper rescaling $V$ by $\xi^{-n}$
and $t$ by $u (\xi^{-n}\q (n-1)!)^{1/n}$, that is
\be
S=V \xi^{-n}, \s \t=t u(\xi^{-n} \q(n-1)!)^{1/n},
\s  \d=\Delta u(\xi^{-n} \q (n-1)!)^{1/n}.
\ee
Now we can rewrite (\ref{equ-1}) as
\be
S(t)=\k{1}{(n-1)!} \int^{\t}_0 \z^{n-1} [1+S(\t-\z-\d)] d\z, \s n=1,2,3,...
\label{equ-2}
\ee
By differentiating the equation twice, we have the following
time-delay equation
\be
\pcd{S}{\t}=1+ S(\t-\d).
\ee
This a delay differential equation, whose explicit solutions are not alway
possible depending on the initial conditions. For a proper initial condition
$S(\t)=0$ for $-\d \le \t \le 0$, the solution can be explicitly obtained by
using the time-forwarding method,
\be
S(\t)=\sumj \k{(\t-j \d)^{nj}}{(nj)!}, \s n=1,2,3,...
\label{equ-3}
\ee
Clearly, for $\t \ne 0$ and $\d \ra 0$, the above solution degenerates into
\be
S(\t)=\sumjj \k{\t^{nj}}{(nj)!},
\ee
given by Moukarzel [3]. For $n=1$, $S(\t)=e^{\t}-1$. For $n=2$,
$S(\t)=\cosh \t -1$. However, when $\d \ne 0$, there is no such simple
expressions. It can be expected that the time-delay parameter $\d$ will
have a strong effect on the evolution behaviour $S(\t)$ of an influence.

\section{Fractal Dimension of Delay Small-World Networks}

The evolutionary solution of $S(\t)$ with time $\t$ and time delay $\d$ gives
the areas affected by the the influence such as fire, infectious diseases and
computer virus. One can expect that fractal dimension of the small-world
networks will give a measure of the network properties as first done by
Newman and Watts (1999) which does not include the effect of time-delay.
Since the velocity of the influence travels at a constant speed $u$, we
can rewrite the solution (\ref{equ-3}) in terms of the distance
$r=\t=At/\xi$ where $A=u (\q (n-1)!)^{1/n}$, we have
\be
S(r)=\sumr \k{(r-j \d)^{nj}}{(nj)!}, \s n=1,2,3,...
\label{equ-4}
\ee
Since $S(r)$ increases as $r$ increases or $S(r) \sim r^{D}$, the
fractal dimension $D$ can be calculated by differential formulae
\be
D=\k{d \log S(r)}{d \log r},
\ee
which can be calculated using (\ref{equ-4}). We now show how the time-delay
affects the fractal dimension $D$ of the small-networks.
For $\d =0$, we  have
\be
S(r)|_{\delta=0}=\sum_{j=1}^{\infty} \k{r^{nj}}{(nj)!}
=\k{1}{n} \sum_{j=0}^{n-1} \exp[r e^{i 2 \pi j/n}] -1,
\ee
so that fractal dimension becomes
\be
D=\k{r \{\k{1}{n} \sum_{j=0}^{n-1} \exp[\k{i 2\pi j}{n}+r e^{i 2\pi j/n}]}{\k{1}{n} \sum_{j=0}^{n-1}
 \exp[\k{i 2\pi j}{n}+r e^{i 2\pi j/n}]-1 }
\ee
For $n=1$, this becomes $D=r e^{r}/(e^{r}-1)$.
Clearly, when $r \ra 0$, the limit of $D$ leads to one. One the other hand,
when $r \ra \infty$, the $r-$term dominates, so we have $D \ra r=At/\xi$.
For $\d \ne 0$, there is no general explicit asymptotic expressions
and numerical evaluations can be easily done.
For small time delay and shorter length scale ($r=A t/\xi \ll 1$),
the fractal dimension remains nearly a constant. For $r \gg 1$, the fractal
dimension increases essentially linearly with the radiu $r$, thus the
dimension of a small-world netword depends on the length scale on which
one looks at it. The dependence of
the fractal dimension $D$ on the lengthscale $\xi$ through $r$ (since
$r$ and $\t$ is rescaled by the length scale $\xi$) and the time-delay
$\d$ suggests the multifractal features of the small-networks.

On the other hand, for a network of finite size, we can change the shortcut
probability $p$ to modulate the effective connectivity and fractal dimension.
For a very small $p \ra 0$, the effective length scale $\xi \ra \infty$, so
any finite-sized network satifies the condition $r \ll 1$, and thus the
fractal dimension $D \ra 1$. As $p$ increases, the length scale $\xi$ reduces,
so that the fractal dimension increases and the network is closely
interconnected with increasing efficiency.

In order to compare the analytical solution with numerical simulations,
we use the simulation method given by Watts and Strogatz (1998) and
Newman and Watts (1999). The numerical simulations for
the one-dimensional $n=1$ and the comparison of fractal dimension with
the analytical solution (\ref{equ-4}) is shown in Figure 1 where
the network size $N=500,000$, $\xi=500$ (or $p=0.002$ and $k=2$) and
$\d=0,1,5,10$. We can see that numerical results (marked with $\circ$,
$\diamond$ etc) are in good agreement with the analytical solution (solid
lines) by using (\ref{equ-4}). As expected, for $r \ll 1$, both numerical
results and analytical fractal dimension approach $n=1$. The time-delay
has a very strong effect on the fractal dimension $D$ of the small-world
networks. For $\d \gg 1$, $D$ is substantially reduced compared with
the one without  time-delay $\d=0$, and $D$ is quite near $n$ for most
of the region. The large time-delay actually makes the network become a
larger world network, and the influence slowly spread  in more
localized areas.

The transition from a very small to higher value of $p$ or very long
to moderate length scale $\xi^*=1/(2 p k n)^{1/n}$ suggests that the
increasing shortcut probability $p$ will greatly increase the
fractal dimension $D$ of the small-world network. This has very
important implications for real world networks.
For a small-world network such as the internet, road networks, and
business partnerships, the introduction of a small fraction of
long-range shortcuts or connections, will essentially
make the network behave more effectively since the effective
scale is becoming smaller. Real networks have always some time delay
that usually makes the network larger than networks without time delay.
Thus there is a trade-off between these two competitive factors and
the proper network design is very important to ensure the efficiency
of real networks such as the internet as well as business partnerships.

In summary, numerical simulations
and analytical analysis for small-world networks with time-delay
show that the time-delay parameter $\d$ has a very strong effect on
the fractal dimension and other properties such as the speed and saturation
time of the delay networks. A small-world network can become
larger if sufficient time-delay in the system response is introduced.
On the other hand, in order to make a large-world network transfer into
a small-world, a slightly higher probability $p$ of long-range
random shortcuts are necessary compared with the one without time-delay
because this essentially reduces the characteristic length scale $\xi=O(1/p)$
quite significantly. Further studies of the dynamics of the delay small-world
networks are now in progress.

{\bf Acknowledgement}: The author would like to thank the referee(s) for
their very helpful comments which have greatly improved the manuscript.

\newpage

\section{References}

\begin{enumerate}

\item D J Watts and S H Strogatz, Collective dynamics of small-world networks,
{\it Nature} (London), {\bf 393}, 440-442 (1998).

\item M E J Newman and D J Watts, Scaling and percolation in the small-world
network model, {\it Phys. Rev.}, E {\bf 60}, 7332-7342 (1999).

\item C F Moukarzel, Spreading and shortest paths in systems with sparse
long-range connections, {\it Phys. Rev.}, E {\bf 60}, R6263-6266 (1999).

\item M E J Newman, C Morre and D J Watts, Mean-field solution of the
small-world network model, {\it Phys. Rev. Lett.}, {\bf 84}, 3201-3204 (2000).

\item A Barrat and M Weigt, On the properties of small-world network models,
       {\it Euro. phys. J.}, B {\bf 13}, 547-560 (2000).

\item M Boots and A Sasaki,
      Small worlds and the evolution of virulence:
      infection occurs locally and at a distance, {\it
      P Roy Soc Lond}, B {\bf 266},1933-1938(1999).

\item B Bollobas, {\it Random graphs}, Academic Press, New York, 1985.

\item S A Pandit and R E Amritkar,
      Characterization and control of small-world networks
      {\it Phys. Rev.}, E {\bf 60}, R1119-1122 (1999).

\item M Barthelemy and L A N Amaral,
      Small-world networks: Evidence for a crossover picture
      {\it Phys. Rev. Lett.},  {\bf 82}, 3180-3183 (1999).

\item D J Watts, {\it Small worlds: The dynamics of networks between order and randomness}, Princeton Univ. Press, 1999.

\end{enumerate}

\begin{figure}
\centerline{\includegraphics[width=5in,height=5in,angle=270]{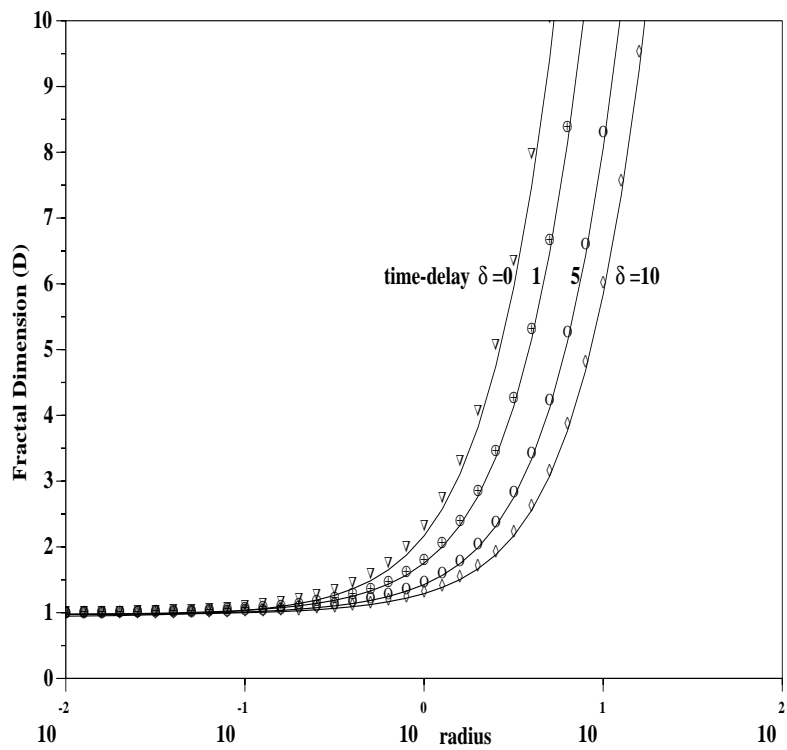}}

\caption{ Fractal dimension of small-world networks with time-delay
$\d=0,1,5,10$ for a network size $N=500,000$, $\xi=500$ (or $p=0.002$)
and $n=1$. Numerical results (marked with $\circ$, $\diamond$ etc)
agree well with analytical express (solid) calculated from (9). }
\end{figure}

\end{document}